# Liquid nitrogen pre-cooling system for ELT instruments utilizing additively manufactured heat exchangers, integrated temperature and liquid level control

Anastasios Aretos[a], Younes Chahid[a], Lee Chapman[a], Mark Cliffe[a], Maia Jones[a], Scott McPhee[a], Chris Miller[a], Graham Wilks[a]

[a] UK Astronomy Technology Centre, Royal Observatory Edinburgh, EH9 3HJ, UK

## ABSTRACT

Using liquid Nitrogen as the cryogenic fluid for pre-cooling purposes is almost ubiquitous across modern observatories. That method tends to be inefficient since it relies on the evaporation of the fluid afforded by the available surface area within the exchanger without capturing any of its latent heat. An additively manufactured heat exchanger could address that inefficiency by providing an increased evaporation area as well as internal structures that can interact with Nitrogen in its gaseous phase. A miniature pre-cooling system utilizing an existing cryostat was developed so the performance of conventionally and additively manufactured heat exchangers would be compared by recording the time required to cool an instrumented mass. Our findings indicate that there is a clear cost and performance advantage of additively manufactured heat exchangers when compared to conventional units, that could lead to reducing the operating costs of observatories hosting large size instruments such as HARMONI.

## 1 INTRODUCTION

### 1.1 Gravity fed pre-cooler -working principle

In its simplest form a pre-cooler can be made by connecting a reservoir holding an amount cryogenic fluid with a heat exchanger and allow gravity to fill the heat exchanger. The subsequent evaporation of the fluid provides the cooling action that reduces the temperature of whatever is attached to the heat exchanger. On Figure 1-1 below a graphical depiction of such a system is provided where the gaseous phase of the cryogenic fluid is removed by the top portion of the pipework attached to the reservoir.

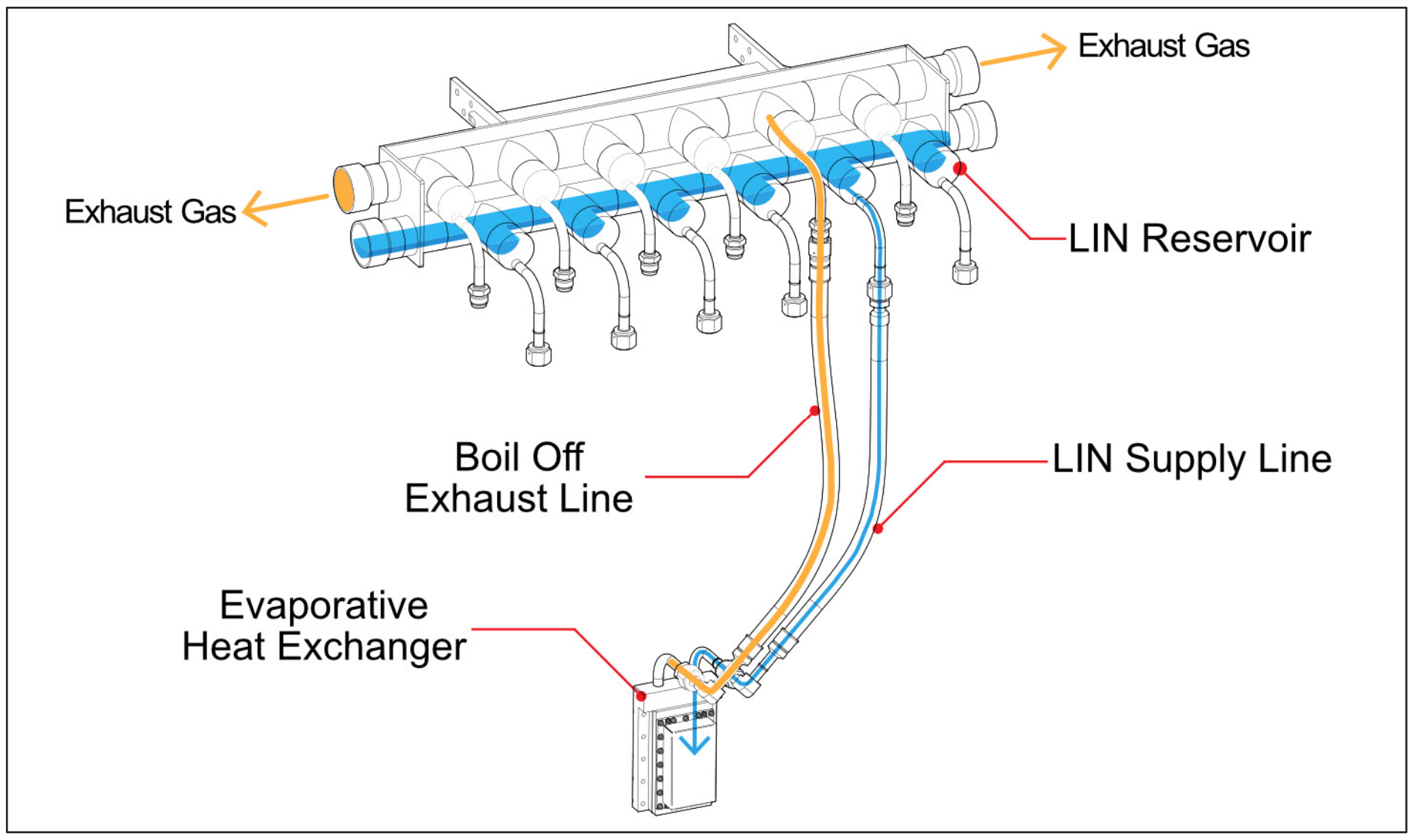

*Figure 1-1 Gravity fed heat exchanger-working principle*

## 1.2 Limitations of a conventional evaporative heat exchanger

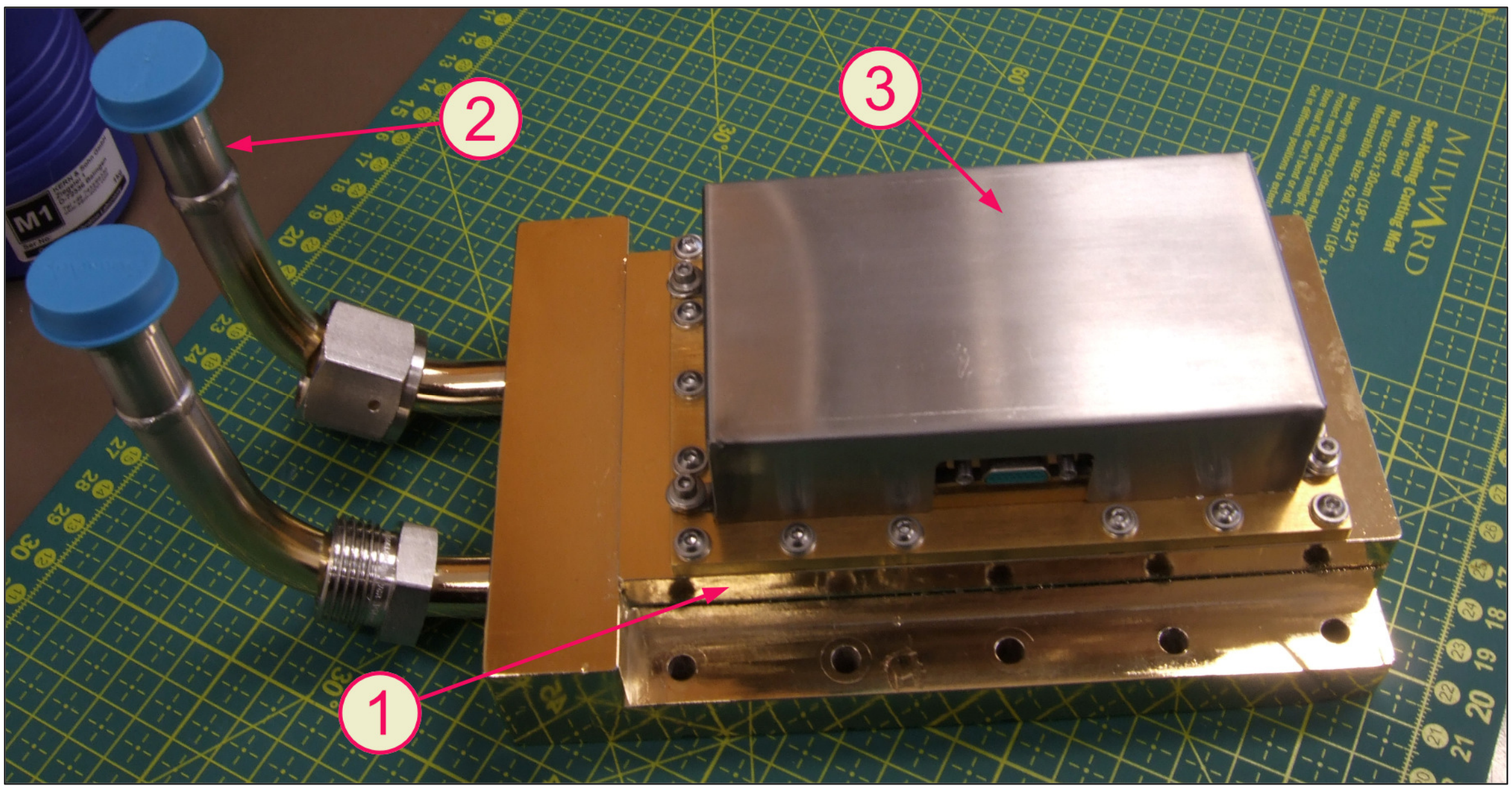

*Figure 1-2 A conventional heat exchanger*

On Figure 1-2 a conventional heat exchanger is shown fully assembled. The main elements of the unit are listed below:

1. Evaporation chamber.
2. Connecting pipework (here furnished with Swagelok VCR™ face sealing nipples).
3. Heater block.

Must be noted that an integrated heater block is not a common element of most heat exchangers. It is included in this unit as the design heralds its origins from heat exchangers developed for instruments developed at the UKATC such as KMOS and MOONS. The example depicted was originally developed for use in the HARMONI instrument.

The evaporation chamber typically contains a set of surfaces that allow for the liquid-to-solid heat exchange to take place. These surfaces can be realized in a variety of ways. A common practice involves machining a pair of solid blocks of metal with good thermal conductivity, such as Oxygen free Copper. The evaporation geometry is machined directly into one of the blocks and the enclosed chamber is created by brazing the other block in situ. Even though brazing is well understood and practiced across many industries, it is a relatively complex multistep manufacturing process. Nonetheless when a heat exchanger needs to come in contact with cryogenic fluids, in this example liquid Nitrogen, whilst is cycled multiple times between cryogenic and room temperatures, achieving a gas tight result becomes quite challenging since brazing material embrittlement, inclusion of brazing impurities in the joint can lead to the heat exchanger developing a permanent gas leak. Consequently, that would make it incompatible with the operation of modern infrared astronomy instruments that contain their optics within vacuum vessels and at cryogenic temperatures.

Furthermore, brazing can only be used on a subset of materials such as steel and Copper and its alloys. In practice, when it comes to cryogenic service components, brazing can be safely employed only when materials with similar coefficients of thermal expansion are considered. Therefore, the geometry that can be designed into a heat exchanger is limited by the material selection as well as the joining method when brazing is considered. The work presented herewith attempts to address some of the abovementioned limitations by removing the need to employ brazing as joining method whilst allowing complex geometries to be realized that would be impractical or economically unfeasible to create by employing conventional methods such as lost wax casting or other near net shape methods especially considering low production volumes..

# 2 METHODOLOGY

## 2.1 Test environment control

In order to test the heat exchangers in a consistent manner a cryostat was employed which was used as a vacuum vessel mostly. Within the cryostat a test rig was sited that included an instrumented cold mass and a miniature, liquid Nitrogen (LIN) pre-cooler consisting of a manifold responsible for delivering LIN to the heat exchanger and exhausting the gaseous Nitrogen out of the system. On Figure 2-1 below a simplified pipework and instrumentation diagram (P&ID) is provided where the main components of the test set up are indicated. As shown on the figure liquid Nitrogen was supplied via a storage dewar, the flow was controlled via a solenoid valve that was actuated based on the feedback received via the liquid level sensor integrated into the manifold. The control system was set so it would attempt to maintain an average fill level within the manifold during operation. The valve used could only be controlled between open and closed states therefore the duration of the actuation was used to modulate the supply of LIN.

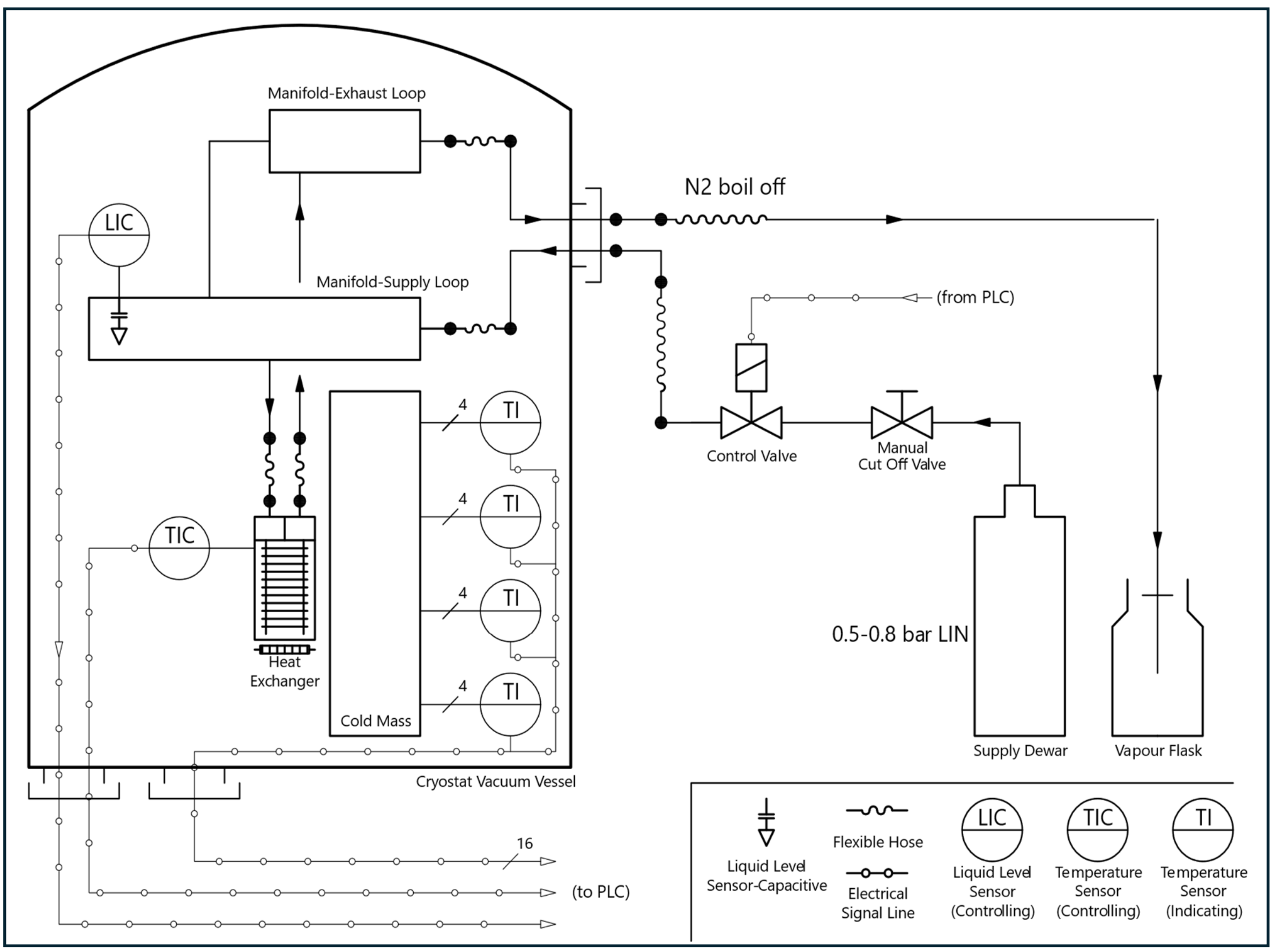

*Figure 2-1 Test set up-pipework and instrumentation diagram*

Apart from the pre-cooler manifold the test rig included a mass of aluminum. That mass was furnished with sixteen PT100 sensors arranged in a grid so the spot temperatures could be collected below and around the attached heat exchanger. Lastly the test rig included a connection to the heater block of the heat exchanger assembly which provided a temperature reading close to the heat exchanger body as well as offered the ability to warm up the test rig prior to backfilling and opening the cryostat. The overall set up was controlled via a pair of MATLAB scripts that controlled the valve on the dewar during cooldown and the heater block during warm up. The tests were conducted in a temperature-controlled room (±2 °C) at UKATC. Although the variance in temperature is not small, its effect was considered negligible because most exposed surfaces within the vacuum vessel were insulated by attaching basic multilayer insulation. The remainder of feed lines and flanges were given foam sleeves to insulate them.

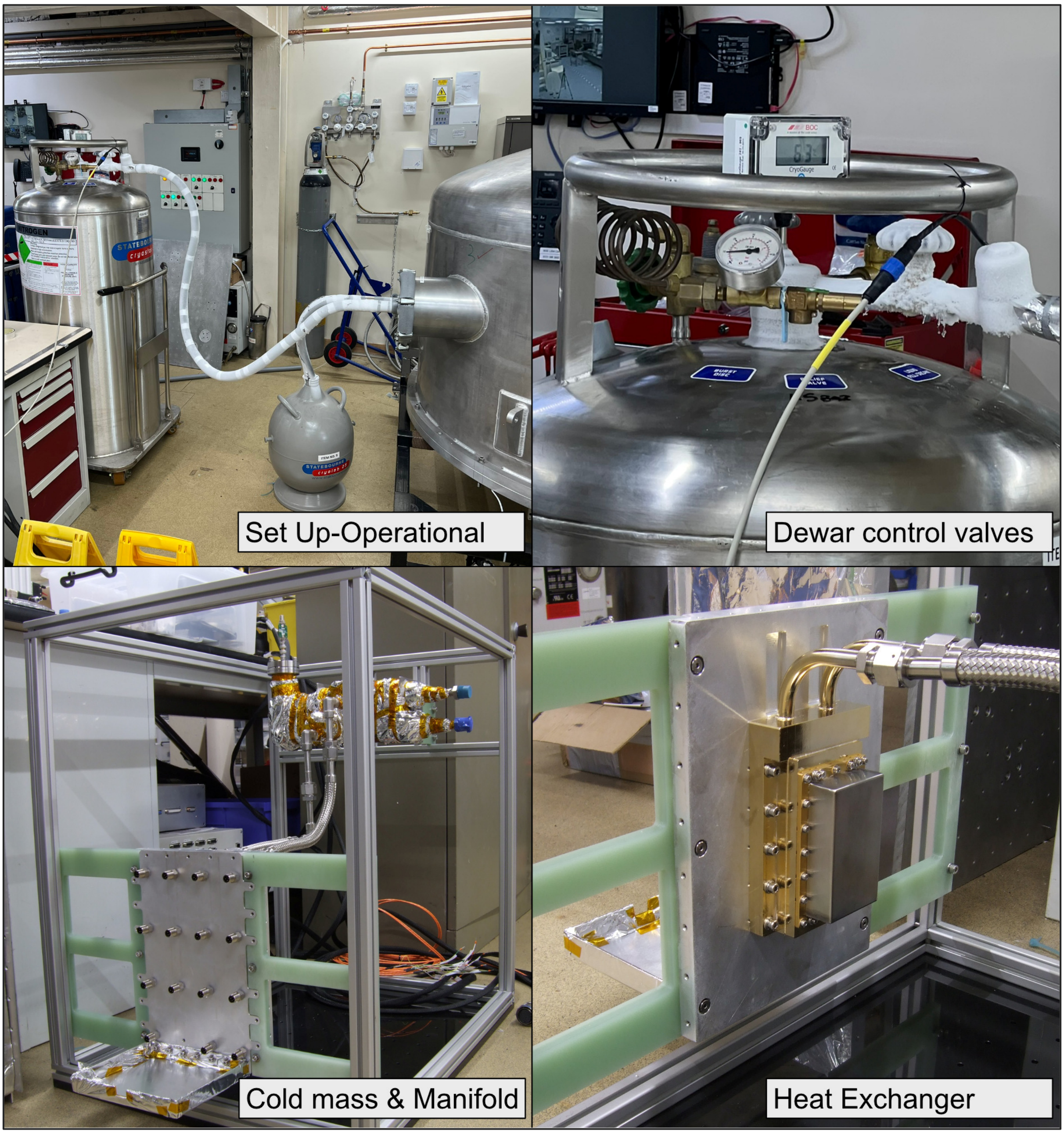

*Figure 2-2 Heat exchanger test set up*

## 2.2 Performance criteria.

Pre-cooling is predominantly applied as a time saving measure in observatories; therefore, the performance of additively manufactured (AM) heat exchangers was evaluated using the total time required to bring the instrumented mass of aluminum to a set temperature. The set temperature and baseline cool down duration was established via the use of a conventionally manufactured unit that was operated on the test rig. A secondary parameter considered was the volume fill value reported by the gauge of the supply dewar in lieu of liquid Nitrogen consumption. Although that is not an ideal measurement given the relatively low sensitivity and resolution of the gauge, it was collected in the absence of readings from a flow meter.

## 2.3 Heat Exchanger Material and Design

Since the conventional heat exchanger was manufactured in Oxygen free Copper (BS EN 13604 CW008A) one material of interested was the pure Copper powder offered by 3D Systems. Conversely, since the main support structures in most large instruments are made in Aluminum Alloys another material of interest was the Scalmalloy™ powder by AP Works, which has already seen some ongoing adoption by METIS [1]. Given the nature of the metal powders the manufacturing method selected was laser powder bed fusion (LPBF), where the part is built in layers by selectively melting portions of a uniformly applied powder layer over a build plate of similar material as the powder.

Another parameter considered was maximizing the evaporation surface whilst maintaining the footprint and mounting of the conventional unit. Specifically, the design explored the relationship between the increase of the evaporation area and the associated reduction of the cool down achieved by the AM heat exchanger made in Copper. To contrast the results obtained by the heat exchanger made in pure Copper powder the heat exchanger made in Scalmalloy™ intended to demonstrate the relative performance difference between the two AM units in relation to the mass reduction and material affinity with the instrumented cold mass of the test rig (BS EN 573-3 AW6061-T6 Aluminum Alloy). Lastly, the AM design used the same volume as the conventional one, to allow for back-to-back comparison and achieved a design balance between a densely increased evaporation area and allowing large enough gaps for powder removal. Other design parameters were the overall cost and ease of integration, evaluated by comparing parameters such as lead time, machining operations required and material purchase cost, between the brazed and welded conventional unit and the AM units that were equipped with double ferrule fittings (Swagelok VCR™ /ISO 8434)

## 2.4 Temperature distribution

With respect to characterizing the overall behavior of the heat exchangers, conventional and additively manufactured, the temperature uniformity of the cold mass was evaluated by generating temperature contour plots by combining the temperature readings from the sixteen PT100 temperature sensors attached to the back of the cold mass (see bottom left inset in Figure 2-2). Although it may be argued that the number of points is relatively small in order to obtain contours that can be considered reliable in depicting the actual temperature distribution of the cold mass the method was deemed adequate given the relativistic nature of the study. On Figure 2-3 below the position of the temperature sensors is shown in relation to the mounted heat exchanger on the opposite side of the cold mass.

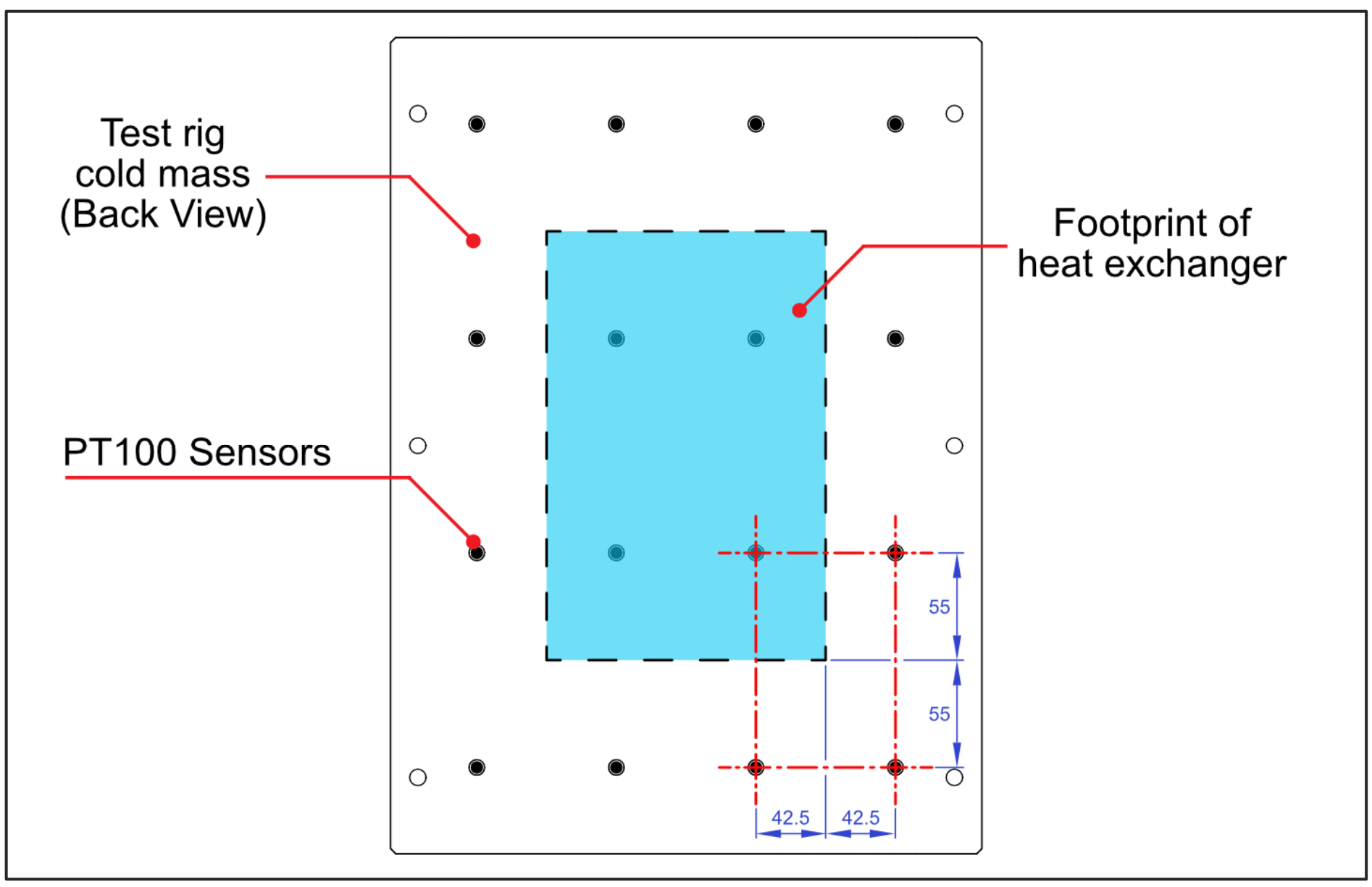

*Figure 2-3 Cold mass temperature sensors arrangement*

## 2.5 Signal Digitizer

The liquid level sensor employed in the test rig is a custom execution of a ¼" unit by Americal Magnetics Ltd. On Figure 2-4 below a basic layout is provided on the left inset. Typically, such sensors require a transmitter (oscillator), so the generated signal can be read and interpreted. The transmitter is usually sited near the connector at the end of the sensor outside the vessel that the sensor monitors [2], the connection is achieved using coaxial cables. A practical limitation of coaxial cables is that their capacitance changes as they shrink due to exposure to low temperatures which makes the reading from the sensor unreliable. This is typically countered by placing the end connection of the monitored tank on the external face of the vacuum vessel, another option employed is the use of tracing heaters that maintain the coaxial cable close to 293 K (20 °C). None of these measures were available in this case, the vacuum vessel could not be modified and a heater traced line would have imposed an excessive heat load.

Exposing a coaxial cable to temperatures as low as 78 K would result in a reported value error up to 100 %, therefore a solution was sought to counter the cable shrinkage effect. The solution developed was to design an analogue to digital signal converter that would be mounted directly onto the sensor. The converter measures the capacitance change of the sensor and digitally signals its value to the control system. As the signal being transferred through the cryostat is digital it remains largely unaffected by the changing capacitance of the cabling. The operating principle of the digitizer is based on charging and discharging a capacitor through a high value reference resistor. That is similarly to the OEM solution offered by American Magnetics. As the level sensor capacitance changes, the period of oscillation of the RC circuit will also change. The change then can be correlated via calibration to the length of the immersed portion of the sensor. The OEM oscillator is not guaranteed to operate in a vacuum environment exposed to cryogenic temperatures. In contrast the digitizer developed is a relatively small board, see top right inset of Figure 2-4, that has proven to work reliably when exposed to cryogenic temperatures and its construction results in very low outgassing, making its adoption possible by current or future instruments.

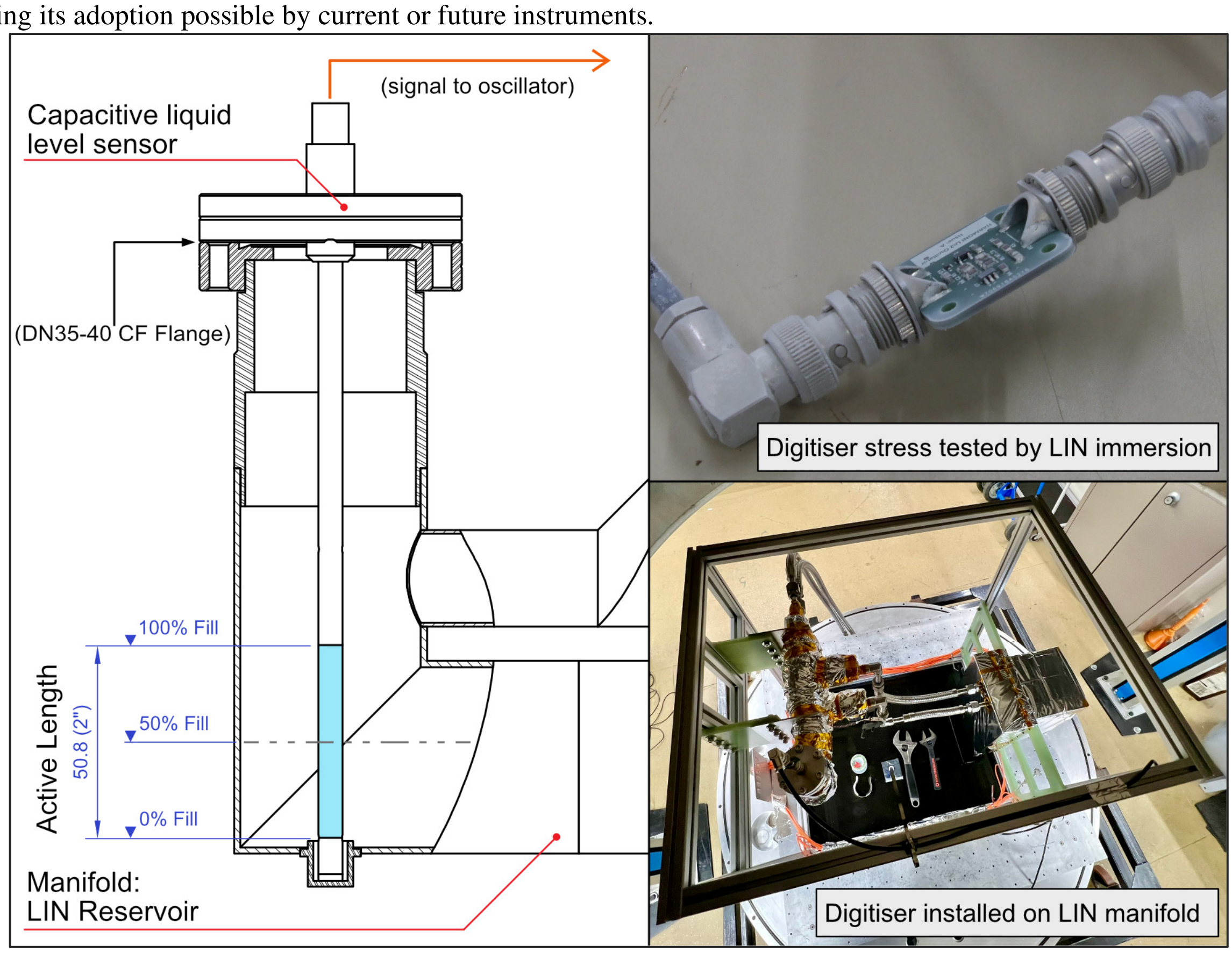

*Figure 2-4 Level sensor-signal digitizer*

### 2.6 Flexible element heater

An integral element of the overall pre-cooler is the addition of a heater block (see Figure 1-2) onto the heat exchanger. Such blocks allow warming up the system prior to backfilling as well as temperature control of specific areas of the geometry. The heater block depicted on Figure 2-5 is constructed by sandwiching a flexible PCB containing a heating coil between two copper plates, on the left inset of the figure a CAD generated image of the assembly is provided highlighting the location of the PT100 sensor and bimetallic thermostat. Typically, the heating element of such heaters is made in materials that have a constant electrical resistance value along a wide temperature range like Constantan. The depicted example was made using Copper as the heating element and a modulated power supply to counter the change to the electrical resistance Copper would demonstrate during operation. The choice of Copper as the heating element was made in order to maintain the cost of the tooling at a manageable level. It can be expected that an operational instance of the heater would be made using Constantan instead if service in an instrument was considered.

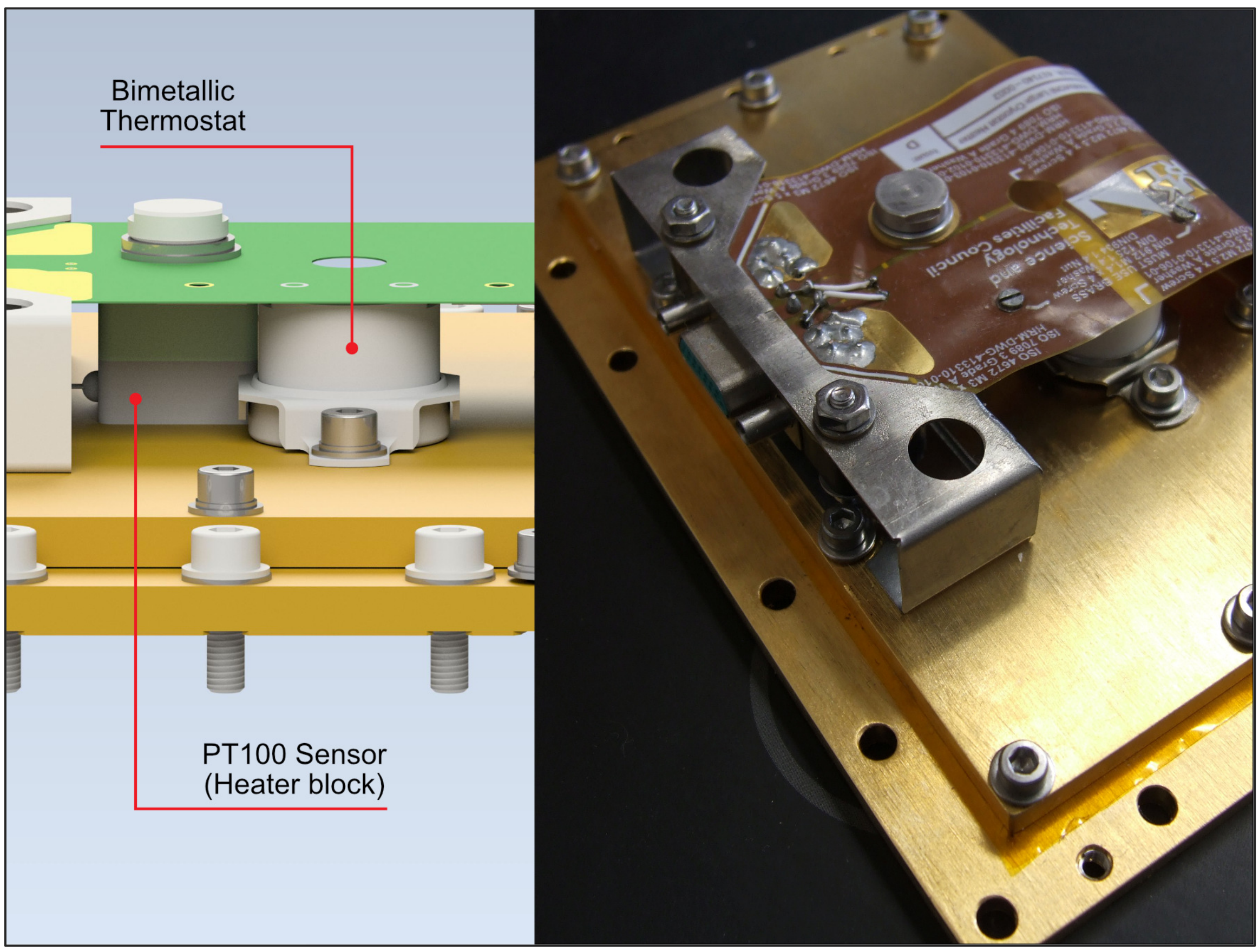

*Figure 2-5 Heater block flexible heater*

As shown on the figure the temperature sensor is located on the top plate that constrains the heating element PCB. The thickness of both plates is 4 mm, and they are made in Oxygen free Copper (BS EN 13604 CW008A) and have been Gold plated to prevent tarnishing of the Copper due to exposure to atmospheric Oxygen and moisture. Although the sitting of the sensor places it, relatively far from the heating element the difference between the reported temperature value and the heating element temperature value is expected to demonstrate a constant offset between the two values therefore making it possible to characterize and calibrate accordingly, if needed.

## 2.7 Additive manufacturing-process development

To reduce the cost of leak testing, benchmark artefacts were additively manufactured across multiple suppliers and materials (pure Copper and Scalmalloy™). Bottom left of Figure 2-6 shows an example of benchmark artefacts that failed the leak test. The chosen suppliers of which the benchmark artefact showed no leak during the hydrostatic testing was 3D systems for pure Copper and Apworks Scalmalloy™. The next stage of development had to do with selecting a gyroid via nTop™ and ANSYS™ to create a triply periodic minimal surface (TPMS) to achieve the desired increase of the evaporation area. The internal TPMS gyroid orientation was adjusted to avoid the use of internal support structures (Top right inset of Figure 2-6). The conventional heat exchanger was designed with 4 mm channels in order to promote the evaporation of Nitrogen that was changed to 5 mm in the case of the AM units, in order to allow for the full evacuation of unused powder out of the internal cavity of the exchanger.

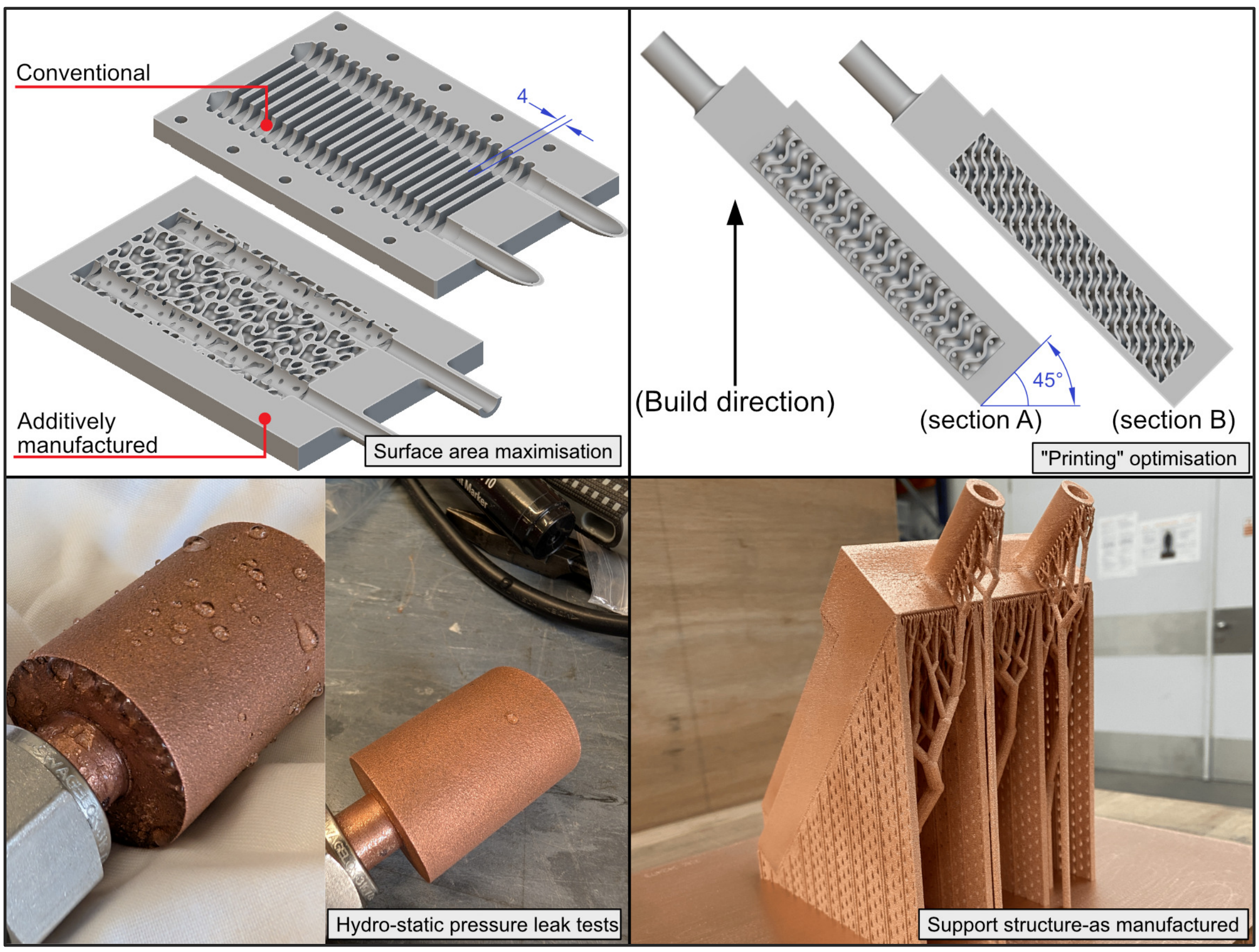

*Figure 2-6 AM Process development steps (Bottom right image credit: 3D Systems)*

Lastly in order to minimize print residual stresses and achieve an easy separation of the part from the build plate the print direction was set to 45° to the vertical so a minimal support structure would need to be generated. In addition, all mounting holes, both through and blind were removed from the geometry, and an extra 2 mm thickness was added on all faces of the part to allow for post-process CNC machining on the components. A view of the support structure is provided on the bottom right inset of the figure. The generation of the printing parameters such as component slicing, etc were all handled by the two respective suppliers, 3D Systems for the exchanger manufactured in Copper and AP Works for the unit manufactured in Scalmalloy™.

# 3 DATA

## 3.1 Key temperature values

As described previously the test set up can collect up to seventeen temperature values across the heat exchanger and instrumented mass. The sample collection rate for every sensor had been set to 1 reading per second, although it was not expected that there would be transient events that would evolve as fast as to require such as high sample rate, it was considered acceptable since the overall file size would still be quite manageable and the extra resolution in the readings could potentially have helped debugging parts of the system. One key temperature value collected is the value reported by the PT100 included into the heater block of the heat exchanger since the block was reused across testing all heat exchangers any errors generated by that part of the circuit were considered as systematic and potentially could be removed via calibration. With respect to evaluating the cooldown performance of each heat exchanger it was decided to monitor one of the 16 sensors on the instrumented cold mass. Since the trial run have indicated that the overall temperature distribution towards the end of the cooldown was minimal across the cold mass the readings from the seventh sensor were used as the criterion, see Figure 3-1 below. All temperatures are reported in K.

Temperature contours were generated by arranging the readings to match the 4×4 grid at the back of the cold mass and passed through MATLAB™ where the shape preserving spline option was utilized in generating the shape of the two-dimensional contours. In this comparative study absolute contour shape accuracy was not considered necessary.

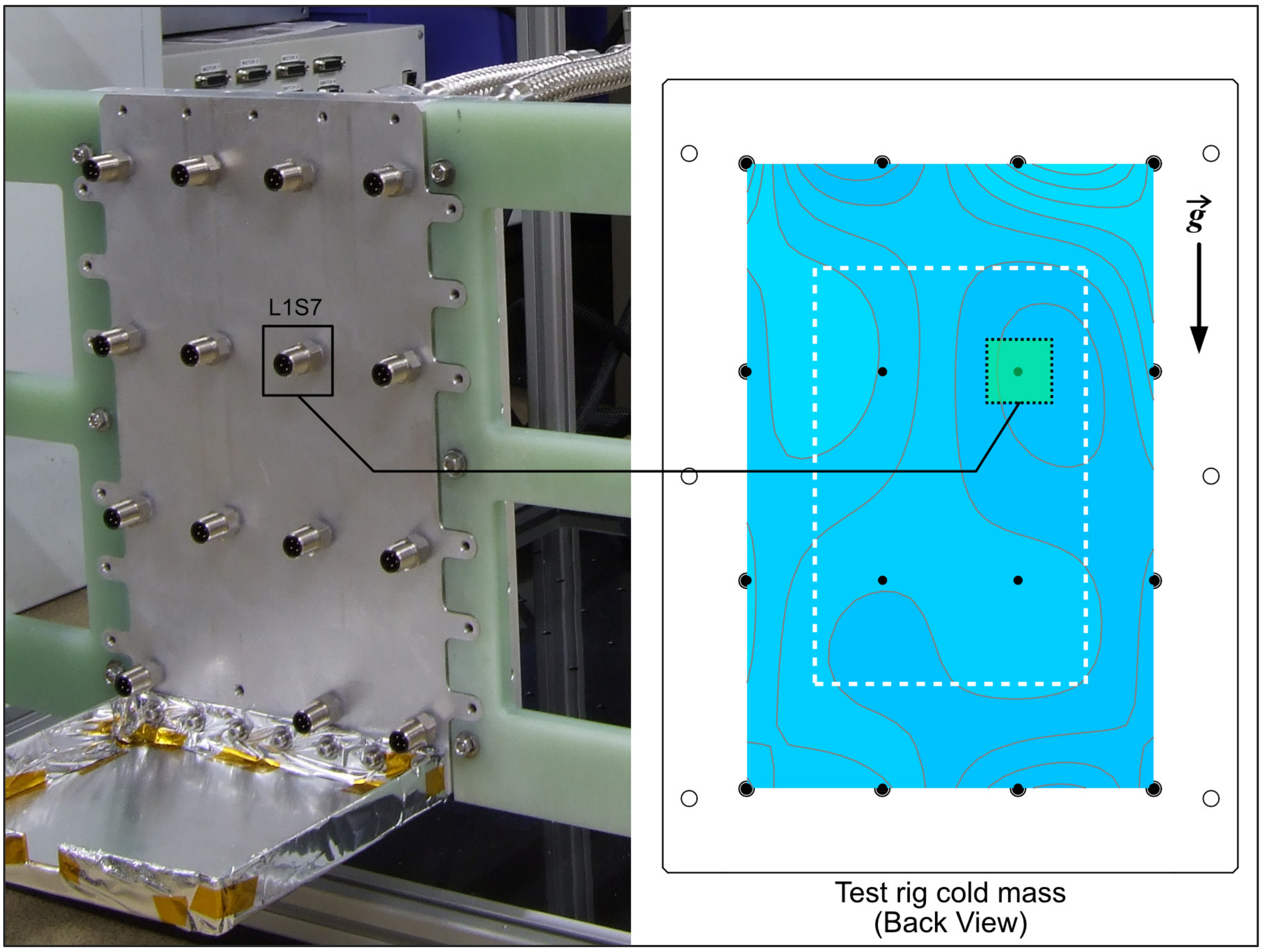

*Figure 3-1 Cold mass temperature probe points*

## 3.2 Liquid Level

The liquid level in the LIN reservoir was reported as percentage of the full calibrated range of the sensor. It was collected to verify that there were not great discrepancies in the LIN delivery by the dewar.

# 4 RESULTS & DISCUSSION

## 4.1 Cool down time

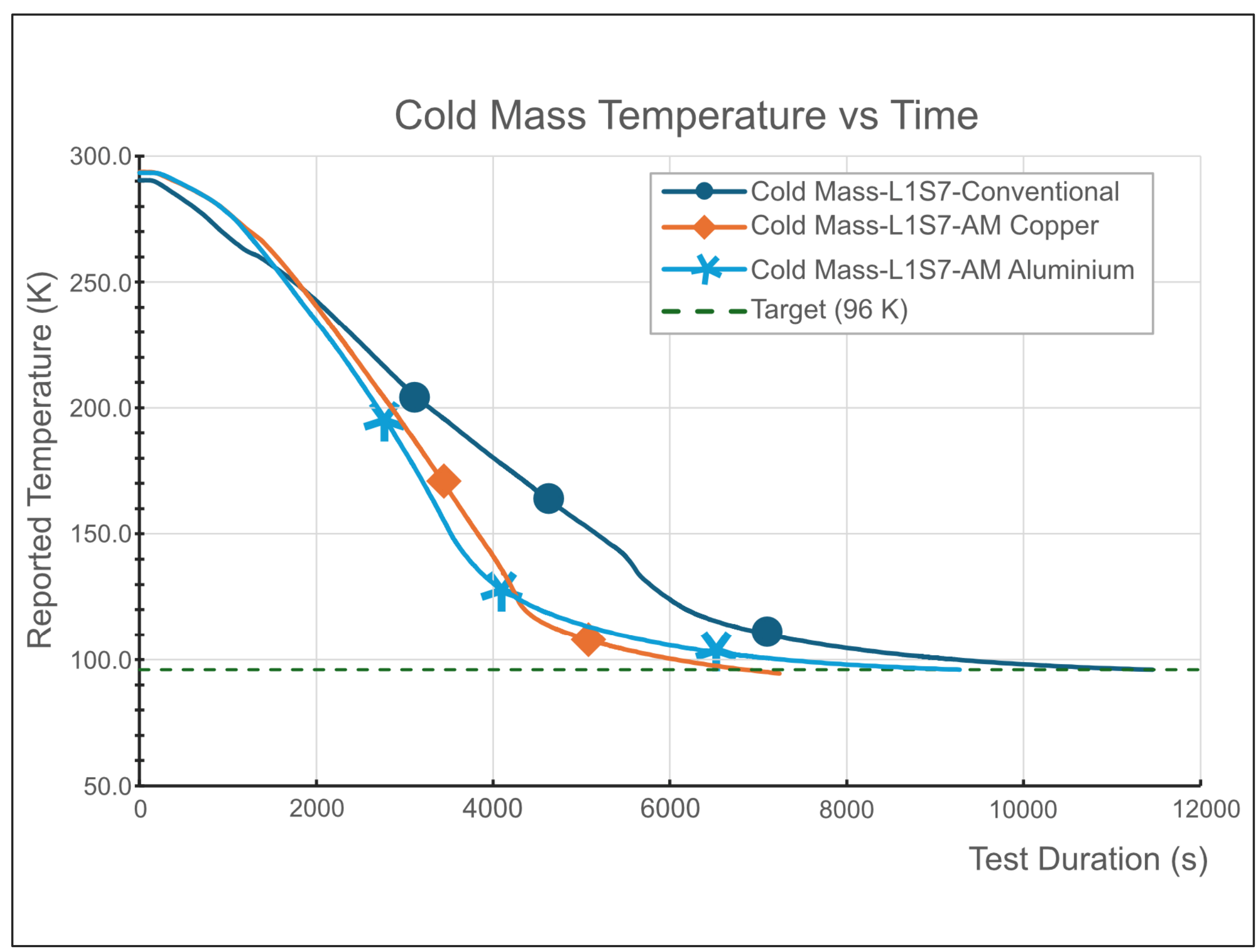

*Figure 4-1 Cold mass cool down time*

On Figure 4-1 above a plot of temperature against time is provided for the cold mass, reporting the value from the sensor highlighted on Figure 3-1. The various results obtained are listed on Table 4-1 below for completeness.

*Table 4-1 Cool down times achieved*

| | **Cold Mass@96 K** | **Heater Block@80 K** |
|---|---|---|
| **Heat Exchanger** | Time to target temperature (s) | Time to target temperature (s) |
| Conventional-CW0008A Cu | 11485 | 6152 |
| AM-Scalmalloy™ (AP Works) | 9277 | 3733.2 |
| AM-Pure Copper (3D Systems) | 6832 | 4332.6 |

With respect to the overall time required to reach the target temperature of 96 K, the AM exchanger in pure Copper powder achieves it within 6832 s. Conversely the reported temperature from the sensor on the heater block indicates that the heater block reaches the temperature of 80 K when attached to the heat exchanger made in Scalmalloy within approximately 3733 s. The result regarding the temperature of the heater block is most likely a direct result of the unit having a smaller overall mass and the base material, Scalmalloy™, having a lower specific heat in comparison to the pure CW0008A Oxygen Free Copper and the pure Copper powder by 3D-Systems.

## 4.2 Temperature distribution

On Figure 4-2 below, a collection of temperature contour plots is provided for the three heat exchangers tested at the end of each test run, after the target temperature was reached (96 K). The overall shape of the contours indicates that there is no clearly discernible effect related to the differential shrinkage between the cold mass and the heat exchanger related to the base material of the heat exchanger as well as the hole mounting pattern of the units. The density of the contours is highest on the plot obtained from the AM heat exchanger in pure Copper powder, and slightly smaller on the plot obtained from the Scalmalloy™ heat exchanger and less on the plot obtained from the conventional heat exchanger. That is probably correlated with the total time required to achieve the target temperature since the cold mass had less time to equalize the faster the heat exchanger reached the target. Nonetheless the total temperature difference bounded by the contours, shown across all three plots, is within 2 K. Therefore, the temperature is reasonably uniform and most likely in an operational setting would be completely equalized since pre-cooling cycles can last several hours based on the experience gained in UKATC integrating instruments such as KMOS and MOONS.

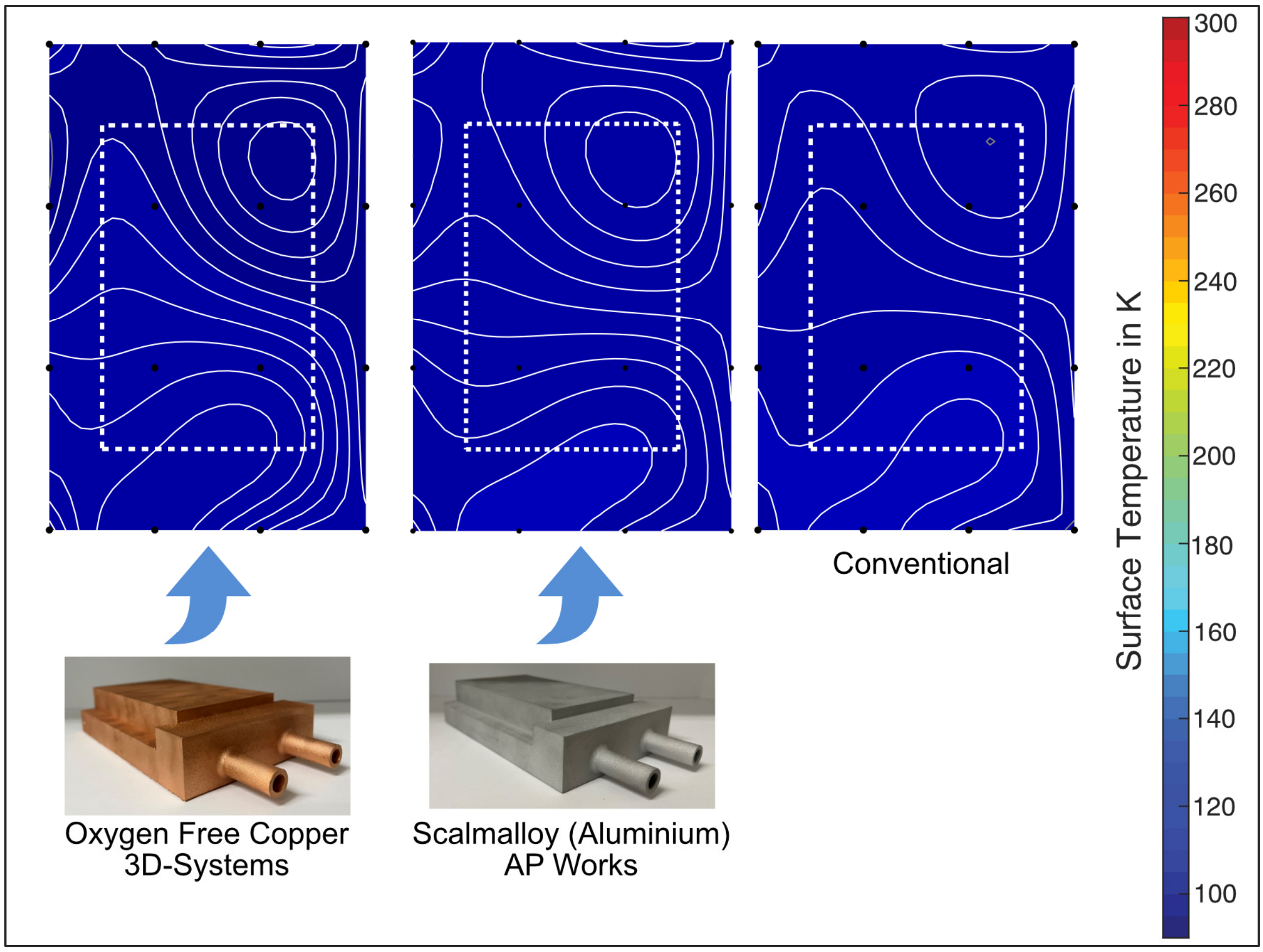

*Figure 4-2 Cold mass-temperature distribution*

The shape of the contours, hints that the flatness of the face on the cold mass may be affecting the overall temperature distribution given the temperature gradient distribution across all three plots. The interface between the cold mass and heat exchangers was not optimized in any way, either by lapping or inclusion of a compliant shim material such as Indium. That was done as to maintain the cold mass face consistent across all tests, whilst the mounting faces across all three heat exchangers were left on their "as machined" state. One difference, though not substantial given the short duration of the test, is that the conventional unit was Gold plated whilst the units made in pure Copper powder and Scalmalloy™ were left in the machined finish state.

### 4.3 Level control and LIN consumption

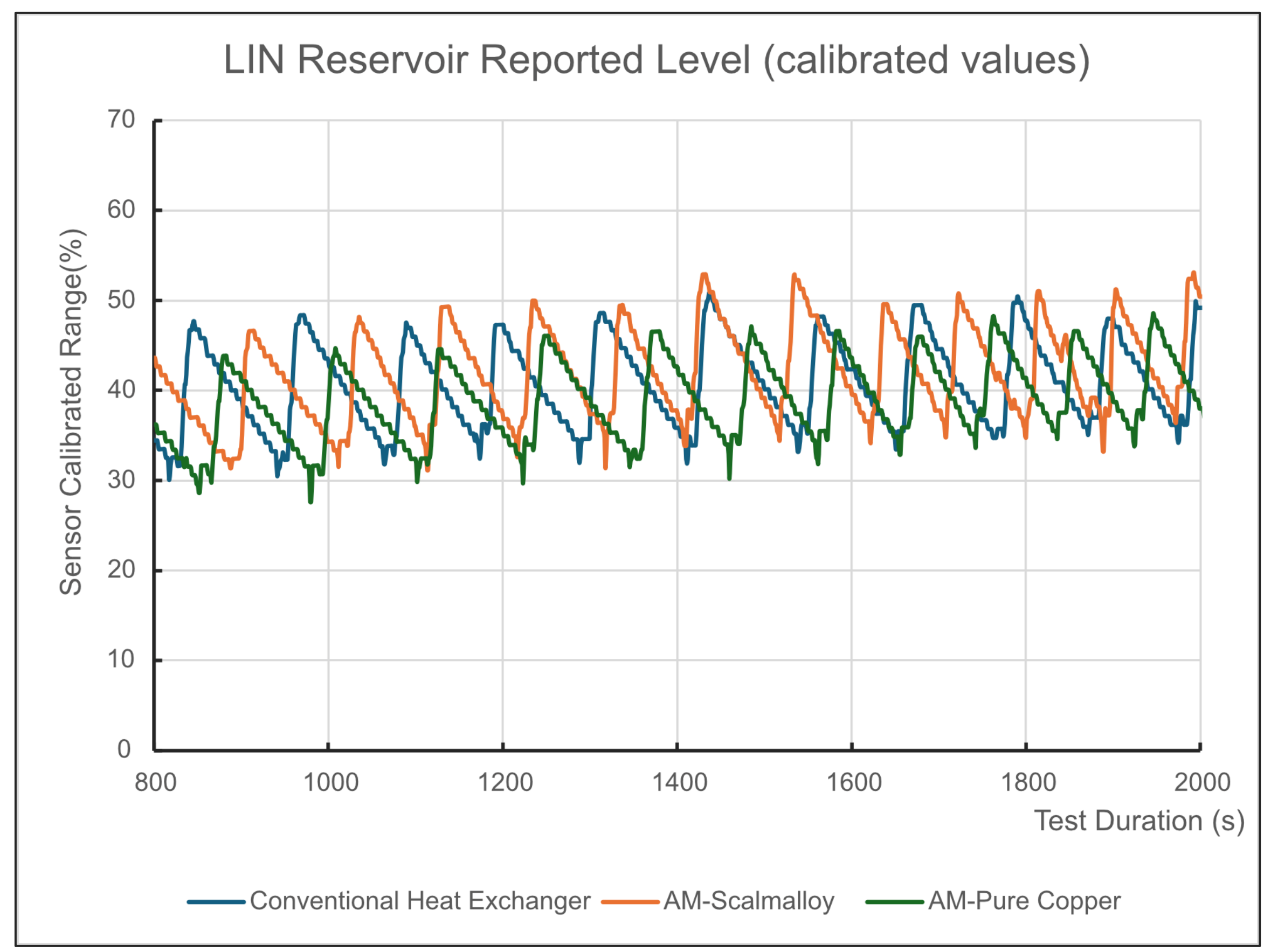

*Figure 4-3 LIN reservoir level*

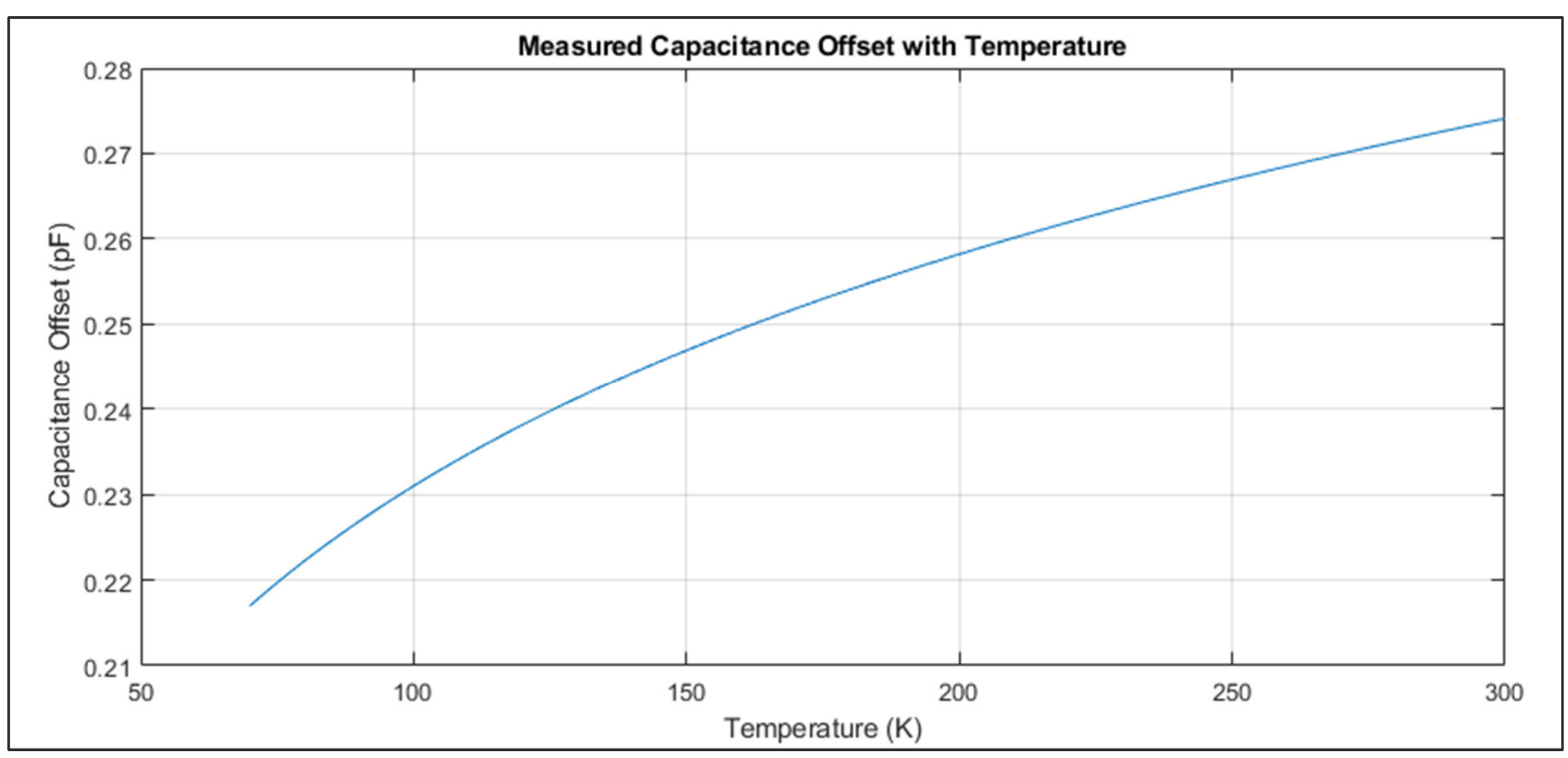

*Figure 4-4 Sensor capacitance offset*

Regarding the performance of the liquid level control portion of the pre-cooler consistent results have been obtained across multiple test runs. On Figure 4-3 above a sample of readings is plotted. The traces correspond to portions of the test runs that generated the temperature results reported on Figure 4-1. In general, the behavior of the system is consistent across all runs. The system responds to a falling liquid level by opening the control valve for a period of time and monitors the level as the manifold discharges LIN to the heat exchanger. The traces indicate that the cut off fill level of the reservoir is not consistent between various actuations. That is a direct result of the magnitude of the temperature dependent parasitic capacitance of the sensor itself. On Figure 4-4 above the value of that capacitance offset is plotted against the temperature of the sensor. The parasitic capacitance originates from the relatively short active length of the sensor. As indicated on Figure 2-4 the active length of the sensor is only 50.8 mm (2”), the overall length of the sensor is 203.2 mm (8”) that makes the measurement of the RC circuit oscillation quite sensitive to small dimensional variations of the sensor stem caused by temperature changes. Typically, that level of noise is negligible since these sensors are made in substantially longer lengths, up to 5.79 m (19 ft). A way to potentially counter that error for such a short sensor would be to employ a near identical sensor mounted on the manifold that is not exposed to the fluid. In that way the control system can remove the temperature dependent offset from the reported level value.

In addition, the liquid level reduction appears to be stepped, whilst the initiation of the reservoir filling is accompanied by a sharp drop and increase in the reported value of the liquid level. Given the proximity of the storage dewar and relatively short length of pipework and absence of a dedicated flow control device up-stream from the manifold, the shape of the traces is most likely the result of liquid sloshing within the manifold. A set of flow baffles realized by attaching perforated plates withing the LIN reservoir probably can resolve that issue.

*Table 4-2 Liquid Nitrogen consumption*

| **Heat Exchanger** | **Dewar fill difference (%)** | **Fill value at the start of run (%)** | **Fill value at the end of run (%)** | **Average room temperature ( °C )** |
|---|---|---|---|---|
| Conventional CW0008A Copper | 17 | 80 | 63 | 20.3 |
| AM-Pure Copper (3D Systems) | 11 | 56 | 45 | 21.6 |
| AM-Scalmalloy (AP Works) | 19 | 44 | 25 | 22 |

The dewar fill percentage was measured at the beginning and end of each run, the results are tabulated on Table 4-2 above. The AM heat exchanger made in pure Copper powder appears to have consumed the least amount of LIN which correlates with the shortest time to reach the target temperature, see Table 4-1 for details. In contrast the largest amount of LIN was consumed by the AM heat exchanger made in Scalmalloy™. Regarding the performance improvement in the consumption of LIN of the AM heat exchanger in pure Copper powder when compared to the conventional unit it can be observed that it is proportional to the time reduction achieved (consumption reduction ≈ 35%, time reduction ≈ 40%). The slightly increased consumption observed in the case of the heat exchanger made in Scalmalloy™, it can be attributed to the reduced specific heat and conductivity of aluminum alloys in comparison to that of Copper. Although reliable values of specific heat and thermal conductivity down to cryogenic temperatures, currently do not exist for the materials used in making the AM units, both materials fall reliably within the general categories of Oxygen free Copper and Aluminum alloys. So, their macroscopic behavior may be extrapolated by the knowledge available on their counterparts. Another factor that may have contributed to the increased LIN consumption could have been the slightly increased room temperature of the laboratory where the test set up was sited, though that could be expected to have a second order effect to the overall behavior of the system given that the room temperature variation between the test runs spans a total of 1.7 °C.

### 4.4 Multi factor performance comparison

An encompassing comparison of the performance obtained from the various heat exchangers does need to include a variety of factors alongside cool-down times or LIN consumption. To that effect on Table 4-3 below the overall lead time and production cost of all the heat exchangers has been included amongst others.

*Table 4-3 Performance factors*

| | **Conventional CW0008A Cu** | **Additively Manufactured Scalmalloy (AP Works)** | **Additively Manufactured Pure Copper (3D Systems)** |
|---|---|---|---|
| Surface Area (%) | 590 $mm^2$ (baseline) | 890 $mm^2$ (+51%) | 890 $mm^2$ (+51%) |
| Cool down time (%) | 11485 s (baseline) | 9277 s (-19.2%) | 6832 s (-40.5%) |
| LIN Consumption | 34 lt (baseline) | 38 lt (+11.8%) | 22 lt (-35.3 %) |
| Assembly Mass (kg) | 3.8 | 1.758 (-53.74%) | 4.558 (+19.94%) |
| Lead time (weeks) | 12 | 6 | 6 |
| Cost (normalized)* | considered unity (baseline) | Undisclosed (-15.5%) | Undisclosed (+57%) |
| * Includes machining, material purchase, fittings and finishing processes. | | | |

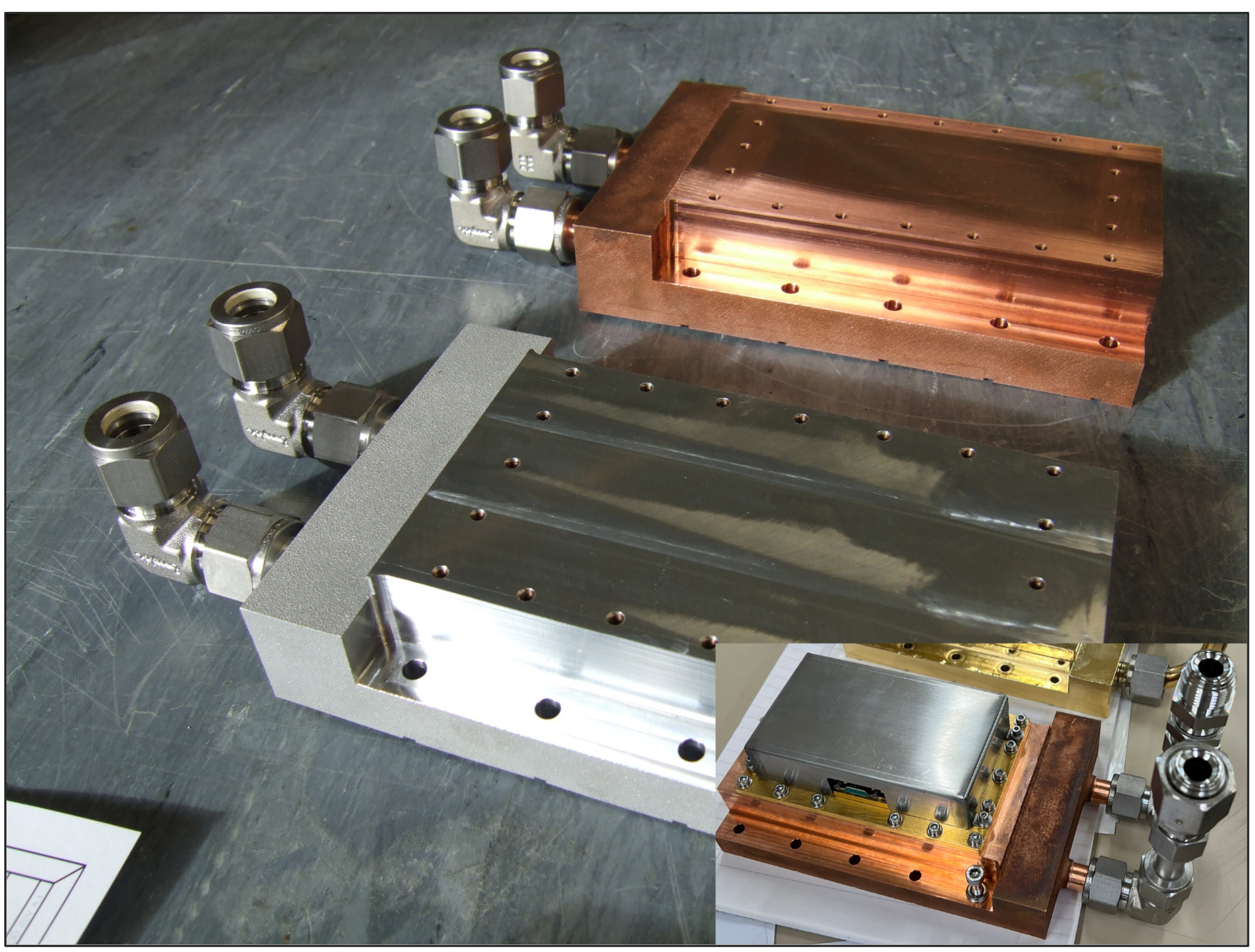

*Figure 4-5 AM Aluiminium and Copper heat exchangers during hardware integration (inset AM pure Copper unit)*

Firstly, given that a common change characterizing both AM units was the increased surface area it would be of value to compare how that change reflects on actual performance improvement.In the case of the unit manufactured in pure Copper powder the overall increase of surface area was a total of 51% which resulted in reduced cool down time by approximately 40% and LIN consumption by 35% whilst the overall mass increased by 20% approximately. Similarly, the unit manufactured in Scalmalloy™ the evaporation surface area was 51% larger in comparison to that of the conventional unit with the cooldown time decreasing by 19% and the LIN consumption increasing by 12% whilst the overall mass was reduced by 53%

Another important set of factors is the lead time and overall production cost to discuss here. Prior to elaborating upon the results obtained the reader is advised to consider that the differences calculated between the conventional unit and the AM manufactured units are particularly skewed due to the extremely low production volumes (one example of each). Furthermore, it must be noted that the manufacturing methods considered so far scale differently when it comes to increased production volumes, therefore the overall differences presented herewith cannot be considered to describe the production cost differences of a larger system containing multiple heat exchangers. Another important aspect to consider here is the monetary value of the production cost. It is presented as a change over a normalized value since any amount would have been valid only at the time of quoting (September 2024) and subject to exchange rate fluctuations as well as it could be considered commercially sensitive, therefore the cost of the conventional unit was used as the baseline and total production cost of the AM units was compared and presented here on a percentage basis only. With all that said, the AM unit in pure Copper powder costs 57% more than the baseline unit, whilst the Scalmalloy™ unit cost 15.5% less. The lower price of the unit produced in Scalmalloy is a direct function of the difference cost between Aluminum and Copper base materials, maybe the difference was compounded by the fact that at the time of production the pure Copper powder was a new material that was introduced by 3D Systems whilst the Scalmalloy™ was an established material on offer by AP Works.

With respect to costs may be noted that the operational cost can be substantially larger than the purchase -production cost of a pre-cooling system especially when larger instruments are considered. That difference can be quite acute if lost observation time is factored against the ability of pre-cooling to rapidly remove heat from an instrument. Therefore, liquid Nitrogen consumption can be considered as indicator of how operating costs may be structured for a larger instrument that could choose to operate conventional or AM heat exchangers. In that regard the unit produced in pure Copper powder offered a reduced cool down time by 40% and LIN consumption by 35% despite the increased production cost by 57% in comparison to the baseline design the performance increase is such that depending on the specific circumstances of an instrument the saving on the operational side of things has the potential to offset the increased production cost. The unit produced in Scalmalloy™ despite the fact it does not offer a similar cooldown time reduction, only 19% reduction vs 40%, and conversely has demonstrated an increased LIN consumption, 11.8% increase, it does offer a substantial mass reduction per heat exchanger, approximately 53%, which could be quite beneficial for systems that are mass limited and the adoption of a copper based heat exchanger may be prohibitive due to the resulting mass.

Lastly must be noted that the AM units did not require any type of brazing or welding operation in order to be made gas tight by consolidating gas fittings inlets/outlets tubes and heater surface component. As shown on Figure 4-5 above, both AM units were post machined and were furnished with double ferrule compression fittings, the ones depicted on the figure belong to the VCR™ product family by Swagelok but any other stainless steel compression fittings to ISO 8434 would have sufficed. As shown on the inset figure, the units were completed by integrating VCR™ face sealing fittings. The integration required basic machining operations such as milling and drilling which were carried out entirely on UKATC's premises. The final result, has proven to be gas tight after exposure to liquid Nitrogen even in the case of the Scalmalloy™ heat exchanger despite the relatively large difference of the coefficient of thermal expansion of the stainless steel and the aluminum alloy comprising the joint. This offers a viable alternative in creating such assemblies without the need for brazing or the use of spring actuated polymer O-rings.

## 5 CONCLUSIONS

1. AM heat exchangers can offer an increased evaporation area within a compact package which translates to a direct improvement of cooling performance both in reduction of cooling times and consumption of cryogenic fluids.
2. AM heat exchangers require fewer manufacturing steps to be realized, and extremely complicated geometries can be realized even when very low production volumes are concerned.
3. The liquid level sensor signal digitizer developed as part of the test set up has performed reliably and consistently during the test campaign. Such a device opens the possibility of sitting liquid Nitrogen reservoirs away from the vacuum vessel walls whilst still maintaining the ability to obtain an accurate liquid level reading.

### ACKNOWLEDGEMENTS

The author would like to thank the staff at the Electronics Workshop of UKATC (Kenny Campbell, Ross Waterston and Arrabella Allison-Donelly) for the support they provided by building the cable harness and flexible heaters used in the test rig. Furthermore, the author would also like to express his gratitude towards the committee in the Centre of Instrumentation of Science Technology Facilities Council of UK Research and Innovation for funding the work and the hardware built during this project as well as the UK Research and Innovation Future Leader Fellowship MR/T042230/1 for supporting access to nTop software.